\documentclass[a4paper,11pt]{article}
\usepackage{jheppub} 
\def\beq#1\eeq{\begin{align}#1\end{align}}

\title{ Semicontinuity of 4d $\mathcal{N}=2$ spectrum under renormalization group flow}

\author[b,c]{Dan Xie}
\author[a,b,c]{Shing-Tung Yau}

\affiliation[a]{Department of Mathematics, Harvard University, Cambridge, MA 02138, USA}
\affiliation[b]{Center of Mathematical Sciences and Applications, Harvard University, Cambridge, 02138, USA}
\affiliation[c]{Jefferson Physical Laboratory, Harvard University, Cambridge, MA 02138, USA}

\abstract{We study  renormalization group flow of four dimensional $\mathcal{N}=2$ SCFTs defined by isolated  hypersurface three-fold singularities. We define the spectrum of $\mathcal{N}=2$ theory as the 
set of scaling dimensions of the parameters on the Coulomb branch, which include Coulomb branch moduli, mass parameters and 
coupling constants. We prove that the spectrum of those theories is semicontinous under the RG flow on the Coulomb branch using the mathematical result about the singularity spectra under deformation. The semicontinuity behavior of $\mathcal{N}=2$ spectrum implies  $\textbf{a}$ theorem under relevant and Coulomb branch moduli deformation, 
the absence of dangerous irrelevant deformations and can be taken as the necessary condition  for the ending point of a RG flow. This behavior is also true 
for $(c,c)$ ring deformation of two dimensional Landau-Ginzburg model with $(2,2)$ supersymmetry. }

\begin{document} 
\maketitle
\flushbottom
%%%%%%%%%%%%%%%%%%%%%%%%%%%%%%%%%%%%%%%%%%%%%%%%%%%%%%%%%%%

\section{Introduction}
The understanding of renormalization group (RG) flow between two conformal field theories is a central subject in study of quantum field theory \cite{wilson1974renormalization}. 
In particular, we would like to know how quantities associated with conformal field theory is changed under RG flow. The most well studied quantity is 
the central charge $a$ in even dimension, and one expect the following $a$ theorem:
\begin{equation}
a_{UV}>a_{IR}.
\end{equation}
This theorem has been studied for 2d theory in \cite{zamolodchikov1986irreversibility}, for  4d theory \cite{cardy1988there,osborn1989derivation,Jack:1990eb,komargodski2011renormalization}, and for 6d supersymmetric field theory in \cite{Cordova:2015fha}. 

A natural further question is that  can we say something about other physical quantities along the RG flow?  Gukov is trying to study how the number of 
relevant operators is changed under RG flow motivated by Morse theory \cite{Gukov:2015qea}, and he proposed a so-called $\mu$ theorem. There
are several directions that one could extend such line of study, for example one might want to study how the scaling dimension of an operator changes under RG flow and 
one should also include operators whose expectation value parameterizes the moduli space of vacua into the study. 

In this paper, we would like to study those RG flow questions for four dimensional $\mathcal{N}=2$ SCFTs. 
The Coulomb branch of $\mathcal{N}=2$ theory 
has a remarkable Seiberg-Witten (SW) solution \cite{Seiberg:1994rs,Seiberg:1994aj}, and we can determine the IR phase exactly. One can also get various interesting SCFTs  by turning on relevant operators and 
the expectation values of Coulomb branch operators. Therefore $\mathcal{N}=2$ theory  is 
a natural place to study RG flows. 
 
Given the importance of Coulomb branch of a $\mathcal{N}=2$ theory, we
define the spectrum of  a $\mathcal{N}=2$ theory as the set of scaling dimensions of parameters on the Coulomb branch. Those parameters include
the Coulomb branch moduli (expectation value of $\mathcal{N}=2$ chiral primary operators), coupling constants, and masses. Let's denote  the rank of charge lattice of a theory ${\cal T}$ as $\mu$, then $\mathcal{N}=2$ spectrum consists a set of  ordered rational numbers: 
\begin{equation}
(\Delta_1,\Delta_2, \ldots, \Delta_\mu),
\end{equation}
and here we have $\Delta_1\geq \Delta_2\ldots \geq \Delta_\mu$.

Let's focus on the theory defined by  isolated three-fold hypersurface singularity \cite{Shapere:1999xr,Xie:2015rpa}. The SW solutions of this class of theories are given by the mini-versal deformation of the singularity,
so the RG flow and its ending point is captured by the deformation of the singularity. 
$\mathcal{N}=2$ spectrum can be computed from the spectra of singularity, which is a set of $\mu$ rational 
numbers $sp(f)=(l_1, l_2, \ldots, l_\mu)$ with ordering $l_1\leq l_2\ldots \leq l_\mu$. The $\mathcal{N}=2$ spectrum can be found using the following simple formula:
\begin{equation}
\Delta_\alpha=1+{1-l_\alpha \over l_1}.
\end{equation}
For each subset $B$ on the real line, we can 
define a non-negative integer counting the number of spectra falling in $B$: 
\begin{equation}
|sp_B(f)|=sp(f)\cap B;
\end{equation}
A subset $B$ is called a semicontinuity domain if 
\begin{equation}
|sp_B(f_{UV})|\geq |sp_B(f_{IR})|,
\end{equation}
for any deformation of $f_{UV}$. The spectrum has the following amazing semicontinuity property which is first conjectured by Arnold \cite{arnold1981some}, and proven in various special cases in \cite{varchenko1982complex,varchenko1983semicontinuity,varchenko1984asymptotics}. The relevant theorem 
for us is the one proven  by Steenbrink \cite{steenbrink1985semicontinuity}:

\textbf{Theorem 1}:  Every half open interval $(t,t+1]$ is a semicontinuity domain of deformations of isolated hypersurface singularity. 

This theorem has various interesting physical consequences. A first consequence is about how the spectrum is changed under RG flow.
Let's assume the ending point of RG flow has the spectrum $(\Delta_{1^{'}},\Delta_{2^{'}}, \ldots, \Delta_{\mu^{'}})$,  
then using above theorem we can show  that every  RG flow  satisfy the following conditions 
\begin{equation}
\Delta_1\geq \Delta_{1^{'}},~~\Delta_2\geq \Delta_{2^{'}}, \ldots;
\end{equation}
This is a much stronger result than $a$ theorem, and
we can actually use it to prove a theorem. Using above theorem, we can prove the absence of dangerous irrelevant deformations, and also put constraints on the possible ending 
point of a RG flow. 

This paper is organized as follows: Section II gives a detailed discussion about the spectrum of $\mathcal{N}=2$ SCFT; Section III discusses 
the identification of $\mathcal{N}=2$ spectrum and the spectrum of singularity; Section IV uses the theorem on change of singularity spectra 
under deformation to prove semicontinouity of  $\mathcal{N}=2$ spectrum. Finally, a conclusion is given in section V.

\section{Spectrum of $\mathcal{N}=2$ SCFT}
Four dimensional $\mathcal{N}=2$ SCFT has a $SU(2)_R\times U(1)_R$ $R$ symmetry. The representation of $\mathcal{N}=2$ superconformal algebra
has been studied in complete detail in \cite{Dolan:2002zh,Beem:2014zpa,Argyres:2015ffa}.  A generic representation is labeled as $|\Delta, R, r, j, \bar{j}>$, here $\Delta$ is the scaling 
dimension, $R$ is $SU(2)$ representation, $r$ is $U(1)$ charge, and $(j,\bar{j})$ are left and right spin. 
There are two sets of special short representations which is of crucial importance for us: ${\cal E}_{r,(0,0)}$ and $\hat{{\cal B}}_R$ (we use the notation of \cite{Dolan:2002zh}). 
The scaling dimension of an operator ${\cal E}_{r,(0,0)}$ is $\Delta={1\over 2}r$ (the scaling dimension of lowest component of the supermultiplet), and the scaling dimension of 
$\hat{{\cal B}}_R$ is $\Delta=2R$. 

${\cal E}_{r,(0,0)}$ can be separated into three different sets:
 a: Relevant operators if $1<\Delta<2$; b: Exact marginal operator if $\Delta=2$; c: others if $\Delta>2$. 
 For the relevant operators and exact marginal 
 operators, one can add the following deformations to our theory
\begin{equation}
\delta S=\lambda \int d^4\theta {\cal E}_{r,(0,0)}+c.c;
\end{equation}
We can assign a scaling dimension to $\lambda$ so that the total deformation is a marginal deformation:
\begin{equation}
\Delta(\lambda)+\Delta({\cal E}_{r,(0,0)})=2;
\end{equation}

One can form a different type of relevant deformation using $\hat{{\cal B}}_R$:
\begin{equation}
\delta S=m \int d^2\theta \hat{{\cal B}}_1+c.c;
\end{equation}
$ \hat{{\cal B}}_1$ has scaling dimension $2$, and $m$ has scaling dimension one. The above two deformations are the only relevant or exact marginal deformations 
for $\mathcal{N}=2$ SCFT, see \cite{Argyres:2015ffa}.

The Coulomb branch are parameterized by the expectation value of the chiral primary operators ${\cal E}_{r,(0,0)}$, and 
it is conjectured that the chiral ring of those operators are a free generated finite ring \cite{Beem:2014zpa}. The low energy effective field theory is described by 
Seiberg-Witten solution which depends on the coupling constants $\lambda$ and masses $m$, and  expectation value  $u$ of chiral operators ${\cal E}_{r,(0,0)}$. So 
the Seiberg-Witten solution has the following schematic form:
\begin{equation}
F(z,u,\lambda, m)=0;
\end{equation}
Using SW solution, we can determine the IR phase for a fixed UV parameters $(u,\lambda, m)$. 
The Coulomb branch plays a crucial role in studying $\mathcal{N}=2$ theory, i.e. the central 
charge can be purely computed from the structure of Coulomb branch.

Let's use $\mu$ to denote the rank of charge lattice $\mu=2r+f$, with $r$  the number of Coulomb branch moduli (number of independent ${\cal E}_{r,(0,0)}$ operators  in Coulomb branch chiral ring), and $f$  the number of mass
parameters. We define $\mathcal{N}=2$ spectrum as a set of $\mu$ ordered rational numbers: 
\begin{equation}
(\Delta_1, \Delta_2,\ldots, \Delta_\mu);
\end{equation}
with ordering $\Delta_1\geq \Delta_2\ldots \geq \Delta_{\mu}$. Here $\Delta_i$ the scaling dimension of the operators ${\cal E}_{r,(0,0)}$ and 
coupling constants $\lambda_i, m$. For operators ${\cal E}_{r,(0,0)}$ with scaling dimension larger than two, we add negative numbers $\Delta[\lambda]$
such that $\Delta(\lambda)+\Delta[{\cal E}_{r,(0,0)}]=2$, notice that such added $\lambda$ parameters would not change the low energy effective 
theory, and they can be thought of irrelevant deformation. The spectrum has the following obvious properties:
\begin{itemize}
\item The spectrum is symmetric with respect to number 1. This is due to definition.
\item The spectrum does not depend on the exact marginal deformation, and this can be seen from the representation theory of $\mathcal{N}=2$ superconformal algebra \cite{Dolan:2002zh}, 
i.e. the operators ${\cal E}_{r,(0,0)}$ and $\hat{{\cal B}}_R$ could not combine with other operators to form long multiplets. 
Obviously the spectrum does not depend on irrelevant deformations.
\item The spectrum satisfies the following condition
\begin{equation}
\sum \Delta_i=\mu.
\end{equation}
\end{itemize}

\textbf{Example 1}: Let's consider $A_3$ Argyres-Douglas theory \cite{Argyres:1995jj,Argyres:1995xn}, and its Seiberg-Witten curve is $x^2=z^4+u_1z^2+u_2 z+u_3$. The charge lattice has rank three. The scaling dimensions of 
various parameters are $[u_1]={2\over 3},[u_2]={1},[u_3]={4\over 3}$, so the spectrum of this theory is 
\begin{equation}
({4\over 3},1,{2\over 3}).
\end{equation}

\textbf{Example 2}: Let's consider $T_3$ theory \cite{Minahan:1996fg,Gaiotto:2009we}. This theory has six mass parameters and one dimension 3 operators, so the spectrum is 
\begin{equation}
(3,1,1,1,1,1,1,-1). 
\end{equation}

The Higgs branch is parameterized by the expectation value of operators $\hat{{\cal B}}_R$, but now the chiral ring is not free generated. However, we would like to conjecture
that the chiral ring of these operators is a finite generated graded affine ring:
\begin{equation}
{\mathbb{C}[x_1,\ldots, x_n]/\mathbb{I}}.
\end{equation}
It would be nice to have a  way to compute this affine ring for general $\mathcal{N}=2$ theory, see \cite{Gaiotto:2008nz,Maruyoshi:2013hja} for some examples. 
More generally, one can have mixed branch which is a direct sum of a Coulomb component and a Higgs component, see \cite{Xie:2014pua} for the structure of 
moduli space of class ${\cal S}$ theory. 

One can have various interacting SCFTs on Coulomb branch, and it is a rather interesting and  difficult problem to classify the IR SCFTs which can appear
on the Coulomb branch of a given UV SCFT. The IR theory on the Higgs branch is rather simple, which is simply a bunch of free hypermultiplets. So 
all the interesting RG flow happens on the Coulomb branch for $\mathcal{N}=2$ theory.

\section{Spectra of singularity and $\mathcal{N}=2$ SCFT}
A large class of $\mathcal{N}=2$ SCFT can be defined by a three dimensional isolated quasi-homogeneous hypersurface singularity $f:(\mathbb{C}^4,0)\rightarrow (\mathbb{C},0)$ \cite{Shapere:1999xr,Xie:2015rpa}. The $\mathbb{C}^*$ action defined on $f$ is
\begin{equation}
f(\lambda^{q_i}z_i)=\lambda f(z_i);
\end{equation}
To define a 4d SCFT, we require $\sum q_i>1$. The SW geometry for the Coulomb branch can be described by the mini-versal deformation of the singularity:
\begin{equation}
F(z,\lambda)=f(z_0, z_1, z_2, z_3)+\sum_{\alpha=1}^\mu \lambda_{\alpha} \phi^{\alpha}=0;
\end{equation}
Here $\phi_{\alpha}$ is the monomial basis of the Jacobi algebra:
\begin{equation}
J_f=\mathbb{C}[z_0, z_1, z_2, z_3]/({\partial f\over \partial z_0},{\partial f\over \partial z_1},{\partial f\over \partial z_2},{\partial f\over \partial z_3});
\end{equation} 
and its dimension is denoted as $\mu$. The scaling dimension of parameter $\lambda_{\alpha}$ is 
\begin{equation}
\Delta_\alpha=[\lambda_{\alpha}]={1-Q_\alpha\over \sum_{i=0}^3 q_i -1};
\end{equation}
Here $Q_{\alpha}$ is the charge of the monomial $\phi_{\alpha}$ under the $\mathbb{C}^*$ action, and 
those scaling dimensions are exactly the spectrum of the corresponding $\mathcal{N}=2$ theory. 

Using mixed hodge structure \cite{steenbrink1976mixed}, one can define a set of $\mu$ rational numbers for the isolated singularity:
\begin{equation}
(l_1,l_2,\ldots, l_\mu),
\end{equation}
and the ordering is $l_1\leq l_2,\ldots \leq  l_\mu$ (These numbers are also related to eigenvalues of the monodromy group associated with the singularity: $m_i=\exp(2\pi i l_i)$), and they are 
called singularity spectra . For quasi-homogeneous isolated singularity, the spectrum can be computed from the Jacobi algebra $J_f$. For 
a monomial basis $\phi_{\alpha}=z_0^{n_0}z_1^{n_1}z_2^{n_2}z_3^{n_3}$, one associate a spectrum number:
\begin{equation}
l_{\alpha}=\sum_{i=0}^3 n_i q_i+\sum_{i=0}^3 q_i-1;
\label{spectra}
\end{equation}
The lowest number in the spectrum of singularity  is $l_1=\sum_{i=0}^3 q_i-1$ and the maximal number is $l_\mu=4-2\sum_{i=0}^3 q_i+l_1$.  For the singularity defining a 4d $\mathcal{N}=2$ theory, 
the spectra is in the range $(0,2)$. 

It is easy to find out $\mathcal{N}=2$ spectrum from singularity spectrum using the following formula:
\begin{equation}
\Delta_\alpha=1+{{1-l_\alpha}\over l_1};
\label{translate}
\end{equation}
It is interesting to note that $l_\alpha=1$ gives us a mass parameter: $l_\alpha=1\rightarrow \Delta_\alpha=1$.

For general isolated three-fold hypersurface singularity, the spectrum of singularity has the following properties \cite{arnol?d2012singularities}:
\begin{itemize}
\item It is symmetric with respect to 1.
\item The sum of the spectrum is $\sum_{\alpha=1}^\mu l_\alpha=\mu$, which implies $\sum_{\alpha=1}^\mu \Delta_{\alpha}=\mu$.
\end{itemize}
The second property is equivalent to $\sum_{\alpha=1}^\mu \Delta_{\alpha}=\mu$. These two properties agree exactly with those of the $\mathcal{N}=2$ spectrum.

\textbf{Example}: Let's consider $A_3$ AD theory defined by the singularity $f=z_0^2+z_1^2+z_2^2+z_3^4$. The weights are $(q_0,q_1, q_2, q_3)=({1\over 2},{1\over 2},{1\over 2},{1\over 4})$. The
basis of  Jacobi algebra is $(1, z_3 ,z_3^2)$, so the spectra of the singularity is $({3\over 4},1,{5\over 4})$ using formula \ref{spectra}.

\textbf{Remark}: In previous work, people usually studied  isolated singularity with a $C^*$ action to define a 4d $\mathcal{N}=2$ SCFT \cite{Shapere:1999xr,Xie:2015rpa}. 
Here we want to conjecture that a general isolated three-fold singularity should also define a $\mathcal{N}=2$ SCFT (The $C^*$ action 
might be thought of as accidental symmetry for the Landau-Ginzburg model on worldsheet if we use  type IIB string description \cite{Ooguri:1995wj,Giveon:1999zm}.). 
The $\mathcal{N}=2$ spectrum is computed from singularity spectrum using formula $\ref{translate}$, and the necessary condition 
for the existence of a SCFT is $l_1>0$ (This is the condition so that the singularity is rational.).

\section{Semicontinuity of spectrum under renormalization group flow}
The Coulomb branch is parameterized by the complex numbers $\lambda_{\alpha}$, and
$\lambda_\alpha$ can be classified by their scaling dimensions:
\begin{itemize}
\item $\Delta_\alpha>1$: Coulomb branch moduli.
\item $0<\Delta_\alpha<1$: Relevant deformations.
\item $\Delta_\alpha=0$: Exact marginal deformations.
\item $\Delta_\alpha<0$: Irrelevant deformations.
\end{itemize}
At a generic point on the Coulomb branch parameterized by $\lambda_\alpha$, the low energy effective field theory is a $U(1)^r$ abelian gauge theory. 
and the corresponding three-fold $F(z,\lambda)=0$ is smooth. If the three-fold is singular, we get interesting $\mathcal{N}=2$ SCFT in the IR. 

We would like to compare the spectrum of the UV theory defined by the hypersurface singularity $f$ and the IR theory on the Coulomb branch. Since $\mathcal{N}=2$
spectrum is  determined by spectrum of the singularity and the RG flow is triggered by mini-versal deformation, 
the equivalent problem is how the singularity spectrum is changed under the deformation.

Interestingly, there is a remarkable theorem about how the singularity spectrum changes under deformation. This is called semicontinuity of spectrum as first conjectured by Arnold \cite{arnold1981some}. 
The theorem relevant for us is proven by Steenbrink \cite{steenbrink1985semicontinuity}. Let's explain some details about this theorem.

Assume that a polynomial $f$ has  singular points $x_1, x_2,\ldots x_n$, we define its spectrum as the union of the spectra of each singular point $x_i$:
\begin{equation}
sp(f)=\sum_{i=1}^n sp(f(x_i)).
\end{equation}
We define the spectrum in the subset $B$ of real line as the number of spectrum numbers in the region $B$:
\begin{equation}
|sp_B(f)|=sp(f)\cap B,
\end{equation}
$B$ is called a semicontinuity domain if 
\begin{equation}
|sp_B(f_{UV})|\geq |sp_B (f_{IR})|.
\end{equation}
Here $f_{UV}$ is the isolated singularity we start with, and $f_{IR}$ is the polynomial at any point on the Coulomb branch.  Steenbrink \cite{steenbrink1985semicontinuity} proves 
the following theorem:

\textbf{Theorem 1}:  Every half open interval $(t,t+1]$ is a semicontinuity domain of deformations of isolated hypersurface singularity. 

In next subsections, we are going to study various interesting physical consequences of this theorem. 

\subsection{Consequences}

\subsubsection{Semicontinuity of $\mathcal{N}=2$ spectrum}
Let's denote the UV spectrum as $l_1\leq l_2\ldots \leq l_\mu$, and the IR spectrum as $l_{1^{'}} \leq l_{2^{'}}\ldots \leq l_{\mu^{'}}$. Using the theorem 1, we have following 
facts about the change of spectrum under RG flow:

\begin{itemize}
\item The first result is:  $l_1^{'}\geq l_1,~l_2^{'}\geq l_2, \ldots$, etc.

\textbf{Proof}: If $l_1^{'}<l_1$, then we choose a domain $B=(l_1^{'}-1, l_1^{'}]$, then $|sp_B(f_{UV})|=0$ and $|sp_B(f_{IR})|=1$, which violates 
the theorem 1. Similarly, one can prove $l_2^{'}\geq l_2$, etc.

We now translate the semincontinuity of the singularity spectrum into the spectrum of the corresponding $\mathcal{N}=2$ theory.  Using the formula $\ref{translate}$, we find 
that: if we denote the UV spectrum as $\Delta_1\geq \Delta_2\ldots \geq \Delta_\mu$, and the IR spectrum as $\Delta_{1^{'}}\geq \Delta_{2^{'}}\ldots \geq \Delta_{\mu^{'}}$, we have 
\begin{equation}
\Delta_1\geq \Delta_1^{'},~\Delta_1\geq \Delta_2^{'}, \ldots. 
\end{equation}
\item  The second consequence is that  $\mu^{'}\leq \mu$. If $\mu^{'}=\mu$, then the spectrum would be the same. We leave the proof to interested reader. 
Since  $\mu$ constant strata in the SW geometry is described 
by the marginal and irrelevant deformations \cite{arnol?d2012singularities}, this result agrees with what is expected from $\mathcal{N}=2$ superconformal algebra. 

\end{itemize}

\textbf{Example}: Let's consider a simple example to illustrate the above theorem. Consider the flow from $A_3$ AD theory to $A_2$ AD theory. The spectrum of UV theory is $({4\over 3}, 1, {2\over 3})$, 
and the IR spectrum is $({6\over5},{4\over5})$, so clearly the spectrum satisfy the semicontinuity property. 

\subsubsection{a theorem}
For the theory defined by hypersurface isolated singularity, 
the central charge can be computed by the following formula  \cite{Shapere:2008zf}:
\begin{equation}
a={R(A)\over 4}+{R(B)\over6}+{5 r\over 24},~~c={R(B)\over 3}+{r\over 6}.
\label{central}
\end{equation} 
Here  $r$ is the rank of Coulomb branch (operator with scaling dimension bigger than one); $R(A)$ can be computed using the spectrum found from SW curve, and $R(B)$ is related to the number of $A_1$ singularity on Coulomb branch:
 \begin{equation}
 R(A)=\sum_{[u_i>1}([u_i]-1),~~~R(B)={\mu \over 4(\sum q_i-1)}={\mu \Delta_1 \over 4}.
 \end{equation}
 Using theorem 1, we have 
 \begin{equation}
 R(A)_{UV}\geq R(A)_{IR},~~R(B)_{UV}\geq R(B)_{IR},~~r_{UV}\geq r_{IR};
 \end{equation}
and we have the equal sign for all three numbers if and only if the deformations are exact marginal or irrelevant deformation. So we have proven that $a_{UV}>a_{IR}$ for the flow triggered by turning on relevant deformations and Coulomb branch moduli.

\subsubsection{Absence of dangerous irrelevant deformations }
A dangerous irrelevant deformation is defined as a deformation which is irrelevant in the UV, but become relevant in the IR. In our setup, those deformations are defined as deformation with negative
scaling dimension. Notice that from our definition, the operator with scaling dimension larger than two is regarded as \textbf{relevant} as one can turn on expectation value of 
these operators and flow to a different IR theory. Using theorem 1,  a deformation with negative 
scaling dimension can never become relevant in the IR.

\subsubsection{Necessary condition for the ending point of a RG flow}
One would like to ask whether a theory $A$ can appear as the ending of a RG flow starting from theory $B$. A necessary condition for this to happen is that the spectrum of A and B should 
satisfy theorem 1.

\section{Conclusion}
In this paper, we found the semicontinuity behavior of  $\mathcal{N}=2$ spectrum under RG flow on  Coulomb branch for theory defined by isolated hypersurface singularity. The weaker statement that 
the scaling dimension of ordered spectrum is decreasing  under RG flow  can be easily checked for some RG flows of
 class ${\cal S}$ theory \cite{gaiotto2012n,Gaiotto:2009hg}  using the result found in \cite{Xie:2014pua}. It would be interesting to check 
 the stronger statement of semicontinuity and find out the semicontinuity domain for this class of theories.
It would be also interesting to explore how spectrum changes under RG flow for general $\mathcal{N}=2$ theory, and we expect that similar semicontinuity behavior is always true.

 Notice that the Higgs branch spectrum 
do not have the monotonically decreasing property under RG flow on Coulomb branch. For example, one can flow from $A_{2n}$ AD theory to $A_{2n-1}$ AD theory on Coulomb branch. The UV theory has no Higgs branch, but the IR theory do have a Higgs branch. 

The semicontinuity property
 is clearly applied to two dimensional $(2,2)$ SCFT defined by Landau-Ginzburg model \cite{Vafa:1988uu,Lerche:1989uy}.  For a two dimensional $(2,2)$ theory defined by a hypersurface isolated singularity with $C^*$ action 
 $f: (\mathbb{C}^n,0)\rightarrow (\mathbb{C},0)$,
 the $(c,c)$ ring is identified with the Jacobi algebra $J_f$ \cite{Vafa:1988uu}, and one can define its spectrum as the $q$ charges of coupling constants of the following deformation
 \begin{equation}
 \delta S=\int d^2\theta \lambda {\cal O}+c.c.
 \end{equation}
 Here ${\cal O}$ is a $(c,c)$ ring element. We have the following set of rational numbers:
 \begin{equation}
 (q_1,\ldots, q_\mu);~~q_1\geq q_2\ldots\geq q_\mu.
 \end{equation} 
These numbers are related to singularity spectrum as $q_i=1-l_i+l_1$. The RG flow is again related to the deformation of singularity, and we have the semi-continouty of $(2,2)$ spectrum 
under RG flow using theorem 1. The physical consequences are similar, i.e. the proof of a theorem, the absence of dangerous irrelevant operator, etc.

Similar behavior about the change of spectrum is also found for RG flow of some four dimensional $\mathcal{N}=1$ SCFTs. The relevant objects are of course the spectrum of chiral ring elements \cite{Cachazo:2002ry}.
Consider $\mathcal{N}=1$ SQCD in conformal window, the chiral spectrum is the mesons $M_i^j=Q_i\tilde{Q}_j$, baryons $B=Q^N$ and anti-baryons $\tilde{B}=\tilde{Q}^N$. The R charge of these operators are
$R(M_i^j)=2{N_f-N_c\over N_f}$, $R(B)=R(\tilde{B})={N_c(N_f-N_c)\over N_f}$. We can turn on mass parameter to one of the flavor to flow from $(N_c, N_f)$ theory to $(N_c, N_f-1)$ theory, and 
the scaling dimensions of chiral operators all decrease under flow.   
It would be interesting to further study how the chiral spectrum is changed under RG flow for general $\mathcal{N}=1$ theory, and we would like to come to this question in the near future. 

Finally, it is interesting to explore how chiral spectrum changes under RG flow for supersymmetric theories in various dimensions. It is also interesting to look at the change of full spectrum under RG flow.

\section*{Acknowledgements}
We would like to thank S.Gukov, Yifan Wang and Wenbin Yan for helpful discussions.
The work of S.T Yau is supported by  NSF grant  DMS-1159412, NSF grant PHY-
0937443, and NSF grant DMS-0804454.  
The work of DX is supported by Center for Mathematical Sciences and Applications at Harvard University, and in part by the Fundamental Laws Initiative of
the Center for the Fundamental Laws of Nature, Harvard University.
%%%%%%%%%%%%%%%%%%%%%%%%%%%%%%%%%%%%%%%%%%%%%%%%%%%%%%%%%%%%%%%%%%%%%%%%%%%%%%%%%%%%%%%%%%%%%%%%%%%%%%%%%%%

\bibliographystyle{utphys}

\bibliography{PLforRS}

\end{document}